\documentclass{jfm}

\usepackage{graphicx}
\usepackage{epstopdf,epsfig}
\usepackage{newtxtext}
\usepackage{newtxmath}
\usepackage{natbib}
\usepackage{hyperref}
\hypersetup{
    colorlinks = true,
    urlcolor   = blue,
    citecolor  = {blue!85!black},
}

\newcommand{\RomanNumeralCaps}[1]
\linenumbers

\usepackage[OT1]{fontenc}
\usepackage[final]{microtype}
\usepackage[x11names]{xcolor}
\usepackage{dcolumn}
\usepackage{bm}
\usepackage[nameinlink,capitalise]{cleveref}

\DeclareMathAlphabet{\mathsfit}{T1}{\sfdefault}{\mddefault}{\sldefault}
\SetMathAlphabet{\mathsfit}{bold}{T1}{\sfdefault}{\bfdefault}{\sldefault}
\providecommand\bnabla{\boldsymbol{\nabla}}
\providecommand\bcdot{\boldsymbol{\cdot}}
\DeclareMathAlphabet\mathsfbi            {OT1}{cmss}{m}{sl}
\IfFileExists{t1phv.fd}
{\DeclareTextFontCommand\textsfbi{\usefont{T1}{phv}{b}{it}}
	\DeclareMathAlphabet\mathsfbi            {T1}{phv}{b}{it}
}{}
\IfFileExists{ot1phv.fd}
{\DeclareTextFontCommand\textsfbi{\usefont{OT1}{phv}{b}{it}}
	\DeclareMathAlphabet\mathsfbi            {OT1}{phv}{b}{it}
}{}

\title{Jamming of Elastoviscoplastic fluids in Elastic Turbulence}

\author{Christopher S. From\aff{1}
  \corresp{\email{christopher.from@manchester.ac.uk}}, 
  Vedad Dzanic\aff{2}
  \corresp{\email{v2.dzanic@qut.edu.au}},
  Vahid Niasar\aff{1},
  \and Emilie Sauret\aff{2}}

\affiliation{\aff{1}Department of Chemical Engineering, University of Manchester, Manchester M13 9PL, UK
\aff{2}School of Mechanical, Medical, and Process Engineering, Faculty of Engineering, Queensland University of Technology, QLD 4001, Australia}

\begin{document}
\maketitle

\begin{abstract}
Elastoviscoplastic (EVP) fluid flows are driven by a non-trivial interplay between the elastic, viscous, and plastic properties, which under certain conditions can transition the otherwise laminar flow into complex flow instabilities with rich space-time-dependent dynamics.
We discover that under elastic turbulence regimes, EVP fluids undergo dynamic jamming triggered by localised polymer stress deformations that facilitate the formation of solid regions trapped in local low-stress energy wells.
Below the jamming transition $\phi<\phi_J$, the solid volume fraction $\phi$ scales with $\sqrt{Bi}$, where $Bi$ is the Bingham number characterizing the ratio of yield to viscous stresses, in direct agreement with theoretical approximations based on the laminar solution. The onset of this new dynamic jamming transition $\phi\geq\phi_J$ is marked by a clear deviation from the scaling $\phi \sim \sqrt{Bi}$, scaling as $\phi \sim \exp{Bi}$. 
We show that this instability-induced jamming transition---analogous to that in dense suspensions---leads to slow, minimally diffusive, and rigid-like flows with finite deformability, highlighting a novel phase-change in elastic turbulence regimes of complex fluids.
\end{abstract}

\section{Introduction}
\label{sec:Intro}
Complex non-Newtonian fluids are well-known to be subject to unpredictable flow instabilities \citep{Datta_etal_PRF_2021,Steinberg_REVIEW}, a problem whose motivations originated from industry over 60 years ago due to production quality issues in processing operations involving polymeric fluids \citep{Petrie_AIChE_1976}, which has continued to intrigue the scientific and industrial communities since \citep{Dubief_AnnuRevFluidMech2023}.
In particular, elastoviscoplastic fluids (EVP) characterized by their ability to exhibit elastic, viscous, and plastic behaviours depending on applied stress are ubiquitous in various processes in nature (such as lava and landslide flow \citep{Rosti_NatPhys,Jerolmack_NatRevPhys} and certain biological substances \citep{Bertsch_2022}) and industry, including materials such as pastes and gels \citep{Balmforth_AnnuRevFluidMech_2014,Alexandre_RevModPhys_2018}. 
EVP materials, a class of non-Newtonian ``yield stress'' fluids, behave as elastic solids with finite deformations below their yield criteria and above which flow like complex viscoelastic fluids \citep{Bonn_RevModPhys2017,Alexandre_RevModPhys_2018}.
Understanding the dynamics of EVP fluid flows remains an essential area of research that bridges fundamental science and practical applications.
A fascinating and challenging aspect of EVP fluid flows is the occurrence of instabilities driven by a delicate balance between this solid-like and liquid-like behaviour that is highly localised and time-dependent \citep{Alexandre_RevModPhys_2018,Varchanis_2020}.
Flow instabilities in EVP fluids are prominent when subjected to deformation rates near their material yield stress, making their applications unpredictable and difficult to control \citep{Balmforth_AnnuRevFluidMech_2014,Alexandre_RevModPhys_2018}.
Insights into these pave the way for innovations in material design and process optimization, addressing challenges across a wide range of industries and disciplines, from the production of fast-moving consumer goods \citep{Balmforth_AnnuRevFluidMech_2014,Bonn_RevModPhys2017} to emerging technologies, such as 3D printing soft biomaterials \citep{Zhang_Science_2024,Smith_SoftMatter_2024,Bertsch_2022}.

Unlike inertial instabilities that arise in Newtonian fluid flows, such as turbulent flows, viscoelastic fluid flow instabilities arise even without inertial effects in the low Reynolds number regime $Re=\rho V \ell/ \mu \leq 1$ \citep{Groisman_2000}.
Inertialess instabilities can manifest in various forms\citep{Datta_etal_PRF_2021}, including shear banding \citep{Fielding_etal_PRL_2024}, where regions of differing shear rates develop, and elastic instabilities, where the fluid's elastic nature leads to flow irregularities. 
The viscous to elastic effects are measured by the Weissenberg number $Wi=\lambda V/\ell\gg1$, and the viscoelastic to inertial effects are characterised by the elasticity number $El\equiv Wi/Re = \lambda \mu /(\rho\ell^2)\gg1$.
Here, $\lambda$ is the longest polymer relaxation time, the characteristic length scale $\ell$, $\rho$ density, $\mu$ total viscosity, and $V$ average velocity.
The additional property, a yield stress criteria $\sigma_y$, traditionally referred to as a jamming transition in EVP material (i.e., the transition from a jammed solid state to a fluid-like state) \citep{Bonn_RevModPhys2017}, and its ratio to the viscous stress is characterised by the Bingham number $Bi=\sigma_y\ell/(\mu V)$.
Time-dependent EVP flows often lead to nonhomogeneity with complex flow patterns and transitions \citep{DzanicFrom2024,Rosti_NatPhys}. 
More specifically, the elastic property generates an anisotropic stress contribution, which, in extreme cases (i.e., $Re \leq 1$), can transition the flow to a chaotic self-sustaining state, known as elastic turbulence (ET) \citep{Groisman_2000,Steinberg_REVIEW}.
Plasticity on its own (i.e., viscoplasticity where there is negligible elasticity) can dramatically impact fluid flows. Notably, recent work \citep{Rosti_NatPhys} demonstrated that plasticity $Bi>1$ significantly alters both the energy distribution and intermittency of inertial turbulence $Re\gg10^3$, altering Kolmogorov's well-known inertial and dissipative 5/3 scaling exponent to a new scaling exponent of 2.3. 
On the other hand, the study of instabilities in inertialess EVP fluid flows, combining both elastic and plastic behaviours, is limited.
Our recent work on EVP extensional flows in the ET regime \citep{DzanicFrom2024} illustrated that the regions of unyielded material increase ---indicated by a higher solid volume fraction ($\phi$)--- as $Bi$ increases. We found that the impact of plasticity on the dynamic behaviour, including transitions to periodic, aperiodic, and chaotic regimes, is highly dependent on the flow geometry.

\textit{In this work}, we discover that EVP fluid flows transition to a jammed state with $Bi$.
The connection to jamming has not previously been made for EVP fluid flows. This elastic-plastic-induced jamming phenomenon is notably distinct from the traditional intuition of jamming transitions in EVP fluids defined as the material’s yield stress criterion $\sigma_y$ \citep{Alexandre_RevModPhys_2018,Coussot_PhysRevFluids_2019}. 
To gain a better understanding of the nature and impact of this phase transition, we demonstrate the direct analogue features of jamming in inertialess EVP flows in the ET regime.
We show that intermediate $\phi-$regimes approaching jamming from below $\phi\rightarrow\phi_J$ follow a square-root scaling behaviour $\phi\sim\sqrt{Bi}$, reminiscent of scaling behaviour in dense suspensions.
Beyond the jamming transition $\phi>\phi_J$, the scaling behaviour transitions dramatically ---a key signature of a phase transition--- growing exponentially $\phi\sim\exp{Bi}$.

\section{Methods}
We numerically study inertialess EVP instabilities with the well-known Kolmogorov flow problem in a 2D domain with double periodic boundary conditions. 
The dimensionless governing equations of the EVP fluid are given by the incompressible Navier-Stokes equation,
\begin{equation}
     \bnabla \bcdot \mathbf{u} = 0, \quad
     Re\frac{D\mathbf{u}}{D t} = -\bnabla P + {\beta}\mathbf{\Delta}\mathbf{u}+ \bnabla \bcdot \boldsymbol{\sigma} + \mathbf{F}_0,
     \label{eq:1}
\end{equation}
coupled with the polymer stress tensor,
$$
\boldsymbol{\sigma}=\frac{1-\beta}{Wi}\Big(f \mathsfbi{C}-\mathsfbi{I}\Big),
$$ 
described by a space-time dependent conformation tensor ($\mathsfbi{C}$) constitutive equation,
\begin{equation}\label{eq:2}
    \frac{D\mathsfbi{C}}{D t} = \mathsfbi{C}\bcdot\left(\bnabla\mathbf{u}\right)
    +\left(\bnabla\mathbf{u}\right)^{\mathsf{T}} \bcdot \mathsfbi{C}
    - \frac{\mathcal{F}}{Wi}\Big(f \mathsfbi{C} - \mathsfbi{I}\Big).
\end{equation}
The functions $f = (L^2 - \operatorname{tr}\mathsfbi{I})/(L^2 - \mathrm{tr}\mathsfbi{C})$ and $\mathcal{F}(\sigma_v,Bi)$ are constitutive polymer models, namely the finite extensible nonlinear elastic Peterlin (FENE-P)\citep{PETERLIN1961257} and the Saramito yield stress model \citep{Saramito_2007}, for the elastic and plastic non-Newtonian behaviour, respectively. 
$L$ is the maximum polymer extensibility ($L^2 > \mathrm{tr}\mathsfbi{C}$), which we set to $L=50$, $\sigma_v$ is the yield stress, $\mathsfbi{I}$ is the identity tensor ($\operatorname{tr}\mathsfbi{I}=2$), $\mathbf{u}$ is the velocity field, and $\mathbf{F}_0$ is the external driving force (the total force $\mathbf{F}= \mathbf{F}_p + \mathbf{F}_0$, where $\mathbf{F}_p = \bnabla \bcdot \boldsymbol{\sigma}$).
\Cref{eq:1,eq:2} are solved using a symmetric positive-definite conserving numerical solver developed in-house \citep{DzanicFrom2024,DzanicFrom_PRE2022,vedad_jfm,DZANIC2022105280,Dzanic_CompFluid_2022}, comprising of the lattice Boltzmann method coupled with a high-order finite-difference scheme.
Let $\boldsymbol{n}$ be the level of periodicity in each direction, where setting $n_x,n_y>1$ results in $n_x\times n_y$ unit cells. For all simulations, we set $n_x=6$ and $n_y=4$, ensuring unicity is conserved, with $N^2=128^2$ grid points in each unit cell (spatial resolution of $2\pi/N$). 
Numerical regularity is added to \labelcref{eq:2} through an additional term $ \kappa\boldsymbol{\Delta}\mathsfbi{C} $ with a specified artificial diffusivity $\kappa$. While this numerical regularity strategy remains highly debated \citep{YerasiGuptaVincenzi_2023arXiv,DzanicFrom_PRE2022,Couchman_JFM_2024}, it is essential to simulating ET flow regimes due to inherent steep polymer stress gradients \citep{vedad_jfm, gupta_vincenzi_2019}. We minimise $\kappa$ and any associated artefacts by setting the admissible Schmidt number $Sc=\nu_s/\kappa=10^3$ as in \citep{Berti_2008}.
Full details on the employed methodology are reported in the Supplemental Material (SM) Sec.~S1.

\begin{figure}
\centering
\includegraphics[width=1.0\textwidth]{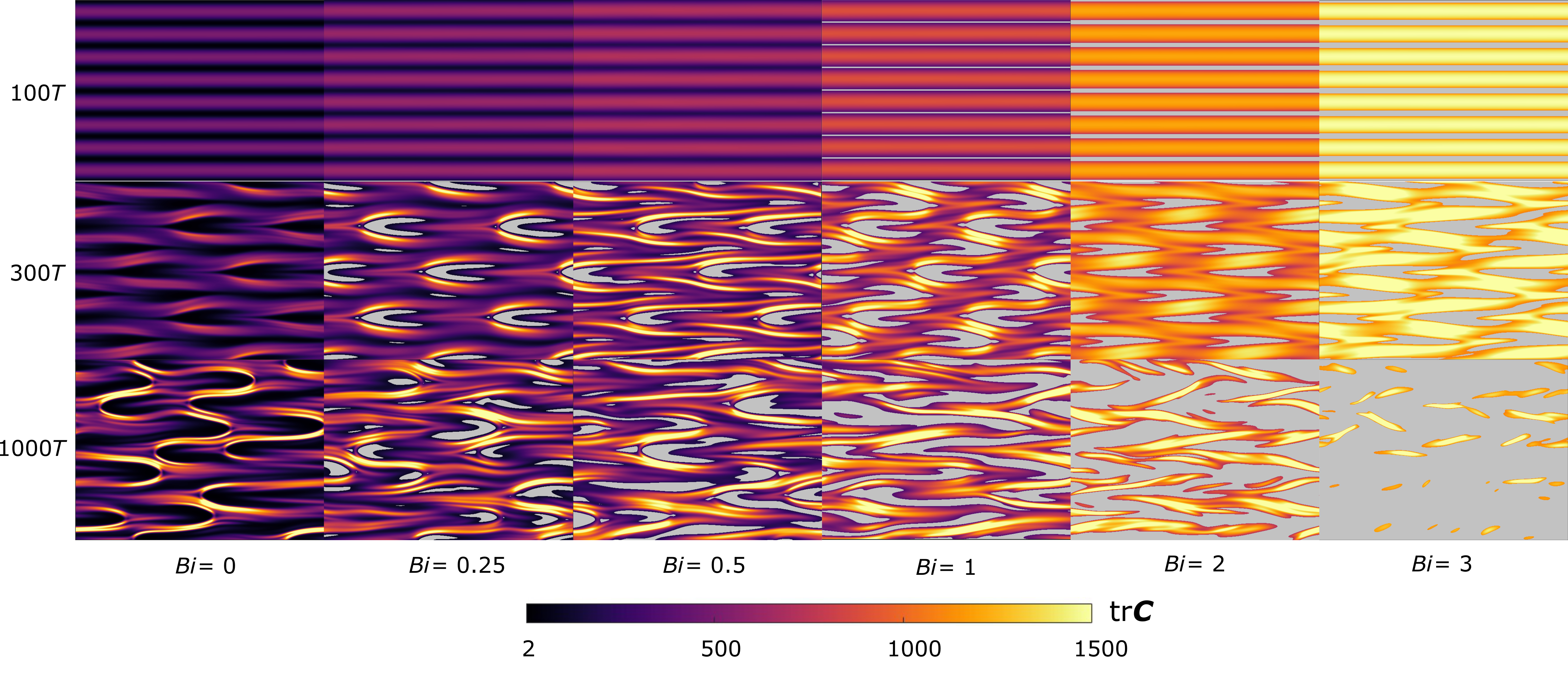}
    \caption{{EVP Kolmogorov flow in the elastic turbulence regime.} Representative snapshots of the polymer stretching $\operatorname{tr}\mathsfbi{C}$ field $x=[0,6\times2\pi)$ and $y=[0,4\times2\pi)$ at various Bingham numbers, from $Bi = 0$ to $ 3$ (columns), at different instances in time (rows): from the initial steady state $\sim100T$ (top) to the transition $\sim300T$ (middle), and the statistically homogenous regime $\overline{t}\gtrsim600T$ (bottom). The grey areas represent the instantaneous unyielded $\mathcal{F}=0$ regions. The first column $Bi = 0$ corresponds to a purely viscoelastic elastic turbulence.}
	\label{fig1}
\end{figure}

The Kolmogorov shear flow is driven by a constant external force $\mathbf{F}_0 = (\mathrm{F}_x,\mathrm{F}_y)$, given by 
$ \mathrm{F}_x(y) = F_0 \cos(Ky), ~ \mathrm{F}_y = 0 $,
with an amplitude $F_0=V\nu K^2$, imposing a constant pressure gradient $\partial_x P = F_0 \cos(Ky)$,
where $K=2$ is the spatial frequency $\ell = 1/K$ and the turnover time $T = \nu_s K/F_0$.
The flow is in the inertialess ET regime, and as such, we set $Re = 1$ and $Wi=20$, such that elastic effects are dominant over inertial effects $El=20$. In doing so, we limit all the observed changes to the well-known viscoelastic Kolmogorov flow \citep{Boffetta_2005,Berti_2008,Berti_2010,Kerswell_JFM2024} to be due to the introduction of plasticity, which we vary by the Bingham number $Bi$. 

\section{Results}
\begin{figure}
\centering
 \includegraphics[width=1\textwidth]{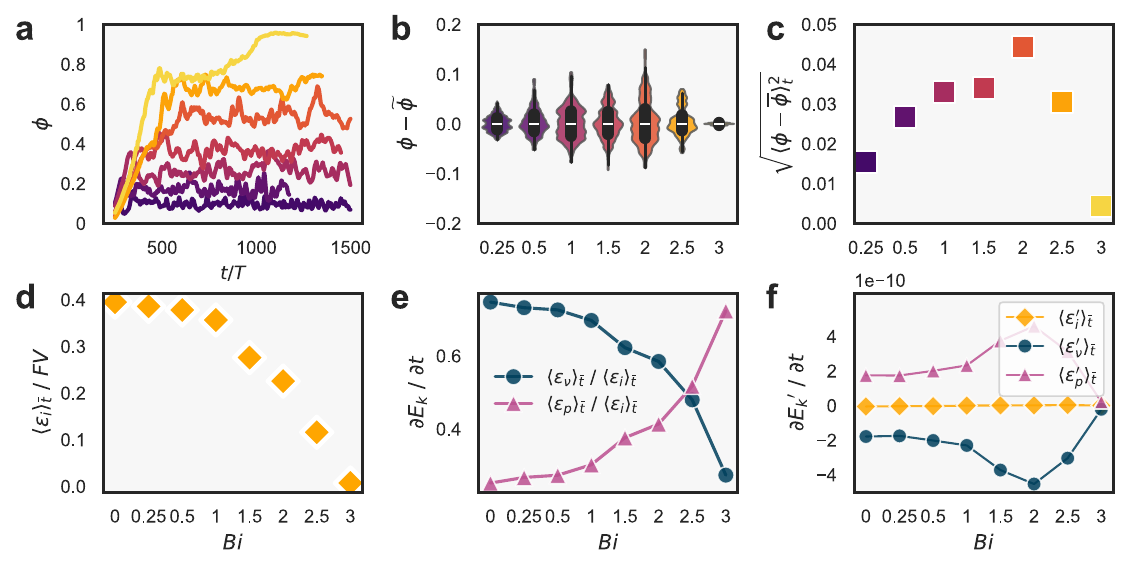}%
        \caption{The solid (unyielded) volume fraction (\textbf{\textsf{a}}) time series $\phi(t)$ and its fluctuation statistics in the statistically homogeneous regime $\overline{t}\gtrsim600T$, including (\textbf{\textsf{b}}) violin distribution density superimposed with box-whisker statics relative to the median $\widetilde{\phi}$ and (\textbf{\textsf{c}}) the root-mean-squared fluctuations, where $\overline{\phi}$ is the temporal mean. 
        The system energy balance of the spatio-temporal mean (\textbf{\textsf{d}}) instantaneous kinetic energy \labelcref{Eq.Energy_Balance}, (\textbf{\textsf{e}}) the viscous dissipation $\epsilon_\nu$ and elastic dissipation $\epsilon_p$, and (\textbf{\textsf{f}}) their corresponding fluctuations \labelcref{Eq.Fluc_Energy_Balance}.
        }
	\label{fig2}
\end{figure}

When dealing with EVP fluids in practice, a macroscopic point of view of the yield transition is usually adopted, where it is assumed that an applied shear above the yield criteria $\sigma_y$ will sufficiently shear-thin the material, allowing it to flow \citep{Dennin_2008,Bonn_RevModPhys2017}. Here, we will model directly how much of the domain remains solid (unyielded) under the imposed flow field. 
The solid volume fraction is calculated by $\phi(t)=v_{\mathcal{F}=0}(t)/v = \lvert\{k:\mathcal{F}(x_k,t)=0\}\rvert / (n_x n_y N^2)$ where ``:'' denotes the set formed by the unyielded regions ($\mathcal{F}=0$) of $k$, and `$\lvert\cdot\rvert$' denotes cardinality \citep{DzanicFrom2024}.
The polymer stretching $\operatorname{tr}\mathsfbi{C}$ field at various $Bi$ is shown in \cref{fig1}. (See SM Movies 1 to 4 corresponding to $Bi=0,~0.5,~2,$ and $3$, respectively.)
The volume fraction $\phi$ time series and statistics are shown in \cref{fig2} (\textbf{a}) and (\textbf{b}--\textbf{c}), respectively. $\phi$ increases with $Bi$ non-trivially, with fluctuations of $\phi(t)$ (yielding and unyielding) in the statistically homogenous region $\Bar{t}$ varying non-monotonically with $Bi$ (\cref{fig2} \textbf{b}--\textbf{c}). 
At $Bi=3$, the flow slowly transitions to a nearly completely jammed state, arresting the flow (as shown \cref{fig2} \textbf{a} and SM~Movie~4).

A notable feature is the formation of coherent structures (CS) of the stress field, commonly referred to as narwhals, which are well-known for the purely viscoelastic case ($Bi=0$) at $Wi>10$ \citep{Boffetta_2005, Berti_2008, Berti_2010,LewyKerswell_arXiv2024} and manifest as travelling elastic waves in the streamwise direction $t\gtrsim300T$ (see, $Bi=0$ case in \cref{fig1} and SM Movie 1).
For EVP cases, similar CS appear due to viscoelastic instabilities of high elasticity, where the addition of spatiotemporal interplay between its viscoelastic-solid and -fluid behaviour leads to further dynamic structural stress deformations of the CS narwhals (see SM Movies).
These modifications to the CS, which manifest even through the introduction of minimal plasticity at $Bi=0.25$ (\cref{fig1}), immediately imply modifications to the modes of elastic waves, altering the transition route to the chaotic self-sustaining ET state \citep{Kerswell_JFM2024,LewyKerswell_arXiv2024}.
The unyielded (viscoelastic solid) regions initially ($t\sim100T$, top row in \cref{fig1}) form between shear layers (i.e., low shear rate regions) before manifesting behind the CS front $t\gtrsim300T$. These low-shear regions effectively act as low-energy wells \citep{Donley_PNAS_2023} that facilitate the formation of unyielded regions where, due to mass conservation, the stress is redistributed between shear layers (see the middle row $\sim300T$ in \cref{fig1}).
During the initial transition $\sim300T$, the length of the unyielded regions in the CS increases with $Bi$ due to a combination of greater $\phi$ (a consequence of increased yield criteria) and increased elastic effects with increasing $Bi$ (at a given constant $Wi$), causing longer streaks. This behaviour is also apparent in the shape of unyielded regions as they become increasingly deformed and elongated.
In the statistically homogeneous regime $\overline{t}\gtrsim600T$ for $Bi=0.25$ to $2.5$, the unyielded regions manifest as local rearrangements with a broad distribution of sizes and shapes leading to further deformations of the CS (see \cref{fig1} and \cref{fig2}~\textbf{b}-\textbf{c}) ---concomitant with the view that plastic events lead to a redistribution of elastic stresses in the system \citep{Alexandre_RevModPhys_2018,Coussot_PhysRevFluids_2019}. 
For $Bi\geq1$, unyielded regions interact across shear layers, merging or splitting each other, where increasing $Bi$ increasingly disorganises the base flow until at the extreme $Bi=3$ where the polymers are stretched in thin and highly localised regions. Notably, $Bi=3$ is the only case with two distinct transitions at $\sim300T$ and $\sim800T$, reaching a statistically homogeneous regime $\overline{t}\gtrsim1000T$ (\cref{fig2}~\textbf{a}).

Increasing $Bi$ decreases the overall energy (\cref{fig2}~\textbf{d}), inferring a shift in the instantaneous kinetic energy balance $\partial E_k/\partial t\approx0$ in the statistically homogeneous regime, 
\begin{equation} \label{Eq.Energy_Balance}
    \frac{\partial E_k}{\partial t} =  \langle\varepsilon_i\rangle_{\Bar{t}}-\langle\varepsilon_{\nu}\rangle_{\Bar{t}} -  \langle\varepsilon_p\rangle_{\Bar{t}} = 0, \quad t\geq\Bar{t},
\end{equation}
where $\varepsilon_i = \langle\mathbf{u}\bcdot\mathbf{F}_0\rangle_{\boldsymbol{x}}$ is the energy input from the Kolmogorov forcing ($\mathbf{F}_0 = (F_0 \cos(y/\ell_y), 0)$), which is dissipated due to contributions from both the viscous Newtonian component $\varepsilon_{\nu}=2\mu_s\langle \mathsfbi{D}:\mathsfbi{D}\rangle_{\boldsymbol{x}}$ and the non-Newtonian polymer component $\varepsilon_p=\langle\mathbf{u}\bcdot(\bnabla\bcdot\boldsymbol{\sigma})\rangle_{\boldsymbol{x}}$. 
Here, $\mathsfbi{D} = \frac{1}{2}\left(\bnabla\mathbf{u} + \bnabla\mathbf{u}^{\mathsf{T}}\right)$ is the rate-of-strain tensor. 
The contributions of $\varepsilon_p$ (elastic and plastic behaviour) increase with $Bi$ and eventually dominate for $Bi\geq2.5$, absorbing more energy than the dissipation of viscous kinetic energy (\cref{fig2} \textbf{e}). 
Fluctuations of the kinetic energy,
\begin{equation} \label{Eq.Fluc_Energy_Balance}
    \frac{\partial {E_k}^{\prime}}{\partial t} =  \langle {\varepsilon_i}^{\prime}\rangle_{\Bar{t}}-\langle{\varepsilon_{\nu}}^{\prime}\rangle_{\Bar{t}} -  \langle{\varepsilon_p}^{\prime}\rangle_{\Bar{t}}= 0, \quad t\geq\Bar{t},
\end{equation}
in which the production term ${\varepsilon_i}^{\prime} =\langle{u_i}^{\prime}{u_j}^{\prime}\frac{\partial\overline{u_i}}{\partial x_j}\rangle_{\boldsymbol{x}} \approx 0$ (\cref{fig2} \textbf{f}) due to negligible inertial ($Re=1$) contributions to the velocity fluctuations ${\mathbf{u}^{\prime}}=\mathbf{u}-\overline{\mathbf{u}}$ \citep{Gotoh1984InstabilityOA,Boffetta_PRE_2005}, where $\overline{\mathbf{u}}=\langle{\mathbf{u}}\rangle_{\Bar{t}}$. 
Instead, the fluctuating non-Newtonian polymer term ${\varepsilon_p}^{\prime}=2\mu_s\langle \mathsfbi{D}^{\prime}:\mathsfbi{D}^{\prime}\rangle_{\boldsymbol{x}}$ is the positive source term to sustain and counteract the fluctuating viscous dissipation ${\varepsilon_{\nu}}^{\prime}=\langle\mathbf{u}^{\prime}\bcdot(\bnabla\bcdot\boldsymbol{\sigma}^{\prime})\rangle_{\boldsymbol{x}}$, which vary non-monotonically with $Bi$ (\cref{fig2} \textbf{f}).
Interestingly, the viscous ${\varepsilon_{\nu}}^{\prime}$ and elastic dissipation ${\varepsilon_p}^{\prime}$ contributions to the energy budget clearly indicate a transition at $Bi=2$, which aligns directly with the strongest and broadest distribution of fluctuations in $\phi(t)$ (\cref{fig2}~\textbf{b} and \textbf{c}). Fluctuations in the dissipation then decrease for $Bi>2$ (\cref{fig2} \textbf{f}) due to the polymer dissipation exceeding the viscous dissipation (\cref{fig2} \textbf{e}), i.e., the system becomes unable to sustain the flow due to strong fluctuations.
The non-monotonic relationship between $Bi$ and the fluctuations $\phi(t)$ along with the impaired flow energy for $Bi\geq2$ are clear features of a typical phase transition ---such a phase transition from a fluid-like to solid-like state is specifically known as \textit{jamming} \citep{Bonn_RevModPhys2017}.

\begin{figure}
\centering
    \includegraphics[width=1\textwidth]{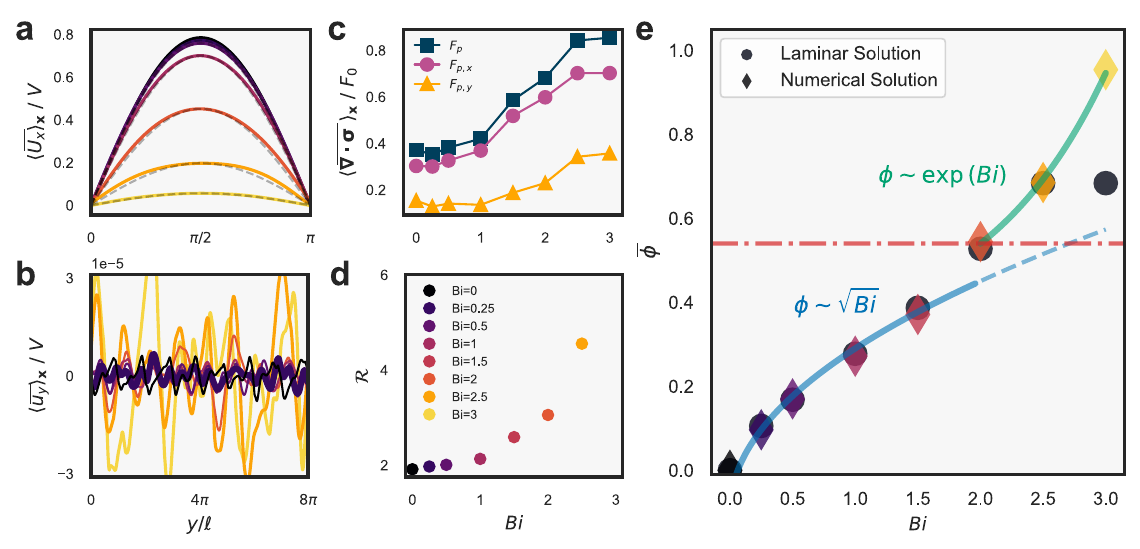}
    \caption{Features of Jamming. Spatiotemporal mean of the  velocity components, (\textbf{\textsf a}) the mean streamise flow profile $\langle \overline{U_x} (y^*)\rangle = (K n_y)^{-1}\sum_{k=0}^{K n_y-1}\lvert \langle \overline{u_x}(y^* + k\pi)\rangle_{{x}} \rvert$ along $y^* \in [0,\pi)$, and (\textbf{\textsf b}) $\langle \overline{u_y} (y) \rangle_{{x}}$, and (\textbf{\textsf c}) the spatiotemporal mean of the polymer force $\langle \boldsymbol{\overline{\nabla\cdot\sigma}} \rangle_{\boldsymbol{x}} = \langle \overline{\mathbf{F}_{p}} \rangle_{\boldsymbol{x}}$. In \textbf{\textsf a}, velocity profiles are superimposed (grey dashed lines) with base sinusoidal profile scaled by the amplitude of each $Bi$ case, i.e., $\max_y(\langle \overline{U_x} (y)\rangle)\operatorname{cos}(Ky)$.
    (\textbf{\textsf d}) Influence of plasticity on the energy injection rate per unit area, measured by flow resistance $\mathcal{R}$ the ratio between the power injected in the statistically homogeneous regime to the base laminar fixed point. 
    (\textbf{\textsf e}) Volume fraction $\phi(t)$ as a function of Bingham number $Bi$, comparing (diamonds) the temporal mean $\bar{\phi}=\langle\phi\rangle_{\overline{t}}$ numerical simulations and (circles) the theoretically approximated $\phi$ \labelcref{eq:laminarSolution} (SM Sec.~S2). For $\phi<\phi_J$, we find that $\phi \sim \sqrt{Bi}$ (blue line) with the linear fit $\phi =0.387 \sqrt{Bi} -9.7\times10^{-2}$ with $\mathrm{R}^2=0.991$. Beyond the jamming transition $\phi_J\simeq0.54$ at $Bi=2$ (red dash-dot line) $\phi \sim \exp{Bi}$ (green line) with the linear fit $\phi =3.2\times10^{-2} \exp{Bi} - 0.24$ with $\mathrm{R}^2=0.999$. 
    }
	\label{fig3}
\end{figure}

A remarkable feature of the Kolmogorov flow is that even in the ET regime, the mean velocity (\cref{fig3} \textbf{a}) and conformation tensor are accurately described by sinusoidal profiles of the base flow with smaller amplitudes with respect to the laminar fixed point \citep{Boffetta_2005,Boffetta_PRE_2005,Berti_2010}. We observe the $\langle \overline{U_x} \rangle$ profile for all $Bi$ retain this feature (see Fig.~S1 in SM) with the velocity front decreasing as $Bi$ increases (\cref{fig3} \textbf{a}). 
This feature is useful since it allows the well-known purely viscoelastic equivalent \citep{Boffetta_2005,Boffetta_PRE_2005,Berti_2010} to be compared. 
For instance, cases for $Bi\leq0.5$ show traces of CS (\cref{fig1}) resembling purely elastic flow behaviour.
The decreasing front velocity amplitude $\langle \overline{U_x} \rangle$ is minor for $Bi\leq1$ but abruptly shifts for $Bi\geq2$ decreasing dramatically, approaching static rest $\langle \overline{U_x} \rangle\rightarrow0$ at $Bi=3$ in \cref{fig3} \textbf{a} (see SM Movie 4). 
This infers a shift in the energy balance, in particular energy dissipation (\cref{fig2} \textbf{d}--\textbf{f}), further evident from the increase in the transverse velocity component $\langle \overline{u_y} (y) \rangle_x$ and the polymer force $\mathbf{F}_{p} = \bnabla\bcdot\boldsymbol{\sigma}$ with increasing $Bi$ in \cref{fig3} \textbf{b} and \textbf{c}, respectively. 
The deformation and locality of plastic events redistribute stresses anisotropically \citep{Alexandre_RevModPhys_2018} as observed in \cref{fig1,fig2}, where the growth of these unyielded regions is strongly correlated with secondary flow effects (\cref{fig3} \textbf{b} and \textbf{c}), leading to a departure from the Kolmogorov base sinusoidal flow profile as $Bi$ increases (\cref{fig3} \textbf{a}). 
Notably, the small transverse loads $\mathrm{F}_{p,y}$ for $Bi>1$ in \cref{fig3} \textbf{c} arise due to finite deformability (\cref{fig1}), which is a well-known feature of traditional jamming in dense suspension flows [see, e.g., \cite{Cates_PRL_1998}]).

We estimate the threshold for the present Kolmogorov flow configuration to be considered jammed $\phi_J\simeq0.54$ at the transition $Bi=2$ (\cref{fig3} \textbf{e}), a value common in many traditional jammed systems, such as dense suspensions \citep{Peters_Nature_2016,Bonn_RevModPhys2017}. 
Approaching the jamming transition from below $\phi\rightarrow\phi_J$ ($Bi<2$), we find that $\phi \sim \sqrt{Bi}$ in \cref{fig3} \textbf{e}. 
Above jamming $\phi\geq\phi_J$ is a clear deviation from $\phi \sim \sqrt{Bi}$, where volume fractions scale as $\phi \sim \exp{Bi}$; such a dramatic change in behaviour further supports the features of phase transition observed in \cref{fig2}.
To theoretically approximate $\phi$ in \cref{fig3} \textbf{e}, we derive the laminar solution (see SM Sec.~S2.1), 
\begin{equation} \label{eq:laminarSolution}
    \mathsfbi{C}_{\mathrm{lam}}(y) =
    \begin{pmatrix}
   f^{-1}(s)\Big(1 + \frac{2 K^2Wi^2}{\mathcal{F}^2 f(s)} \operatorname{sin}^2(Ky)\Big) & K\frac{Wi}{\mathcal{F} f(s)}\operatorname{sin}(Ky) \\
   K\frac{Wi}{\mathcal{F} f(s)}\operatorname{sin}(Ky) & 1
   \end{pmatrix}
   .
\end{equation}
\Cref{fig3} \textbf{e} shows the scaling behaviour, prior to and above $\phi_J$, are in direct agreement with $\phi$ approximated by the laminar solution \labelcref{eq:laminarSolution}. The close agreement is surprising at intermediate $Bi\leq2.5$ given the drastic flow deformations (\cref{fig1}). Interestingly, for $Bi=3$, while \labelcref{eq:laminarSolution} does not predict $\langle\phi\rangle_{\bar{t}}$ in $\overline{t}\gtrsim1000T$ (\cref{fig3} \textbf{e}), it is consistent with $\langle\phi\rangle_{t}$ in the initial steady-state region $300T\lesssim t\lesssim800T$ (\cref{fig2} \textbf{a}).
Analogue scaling behaviour $\phi \sim \sqrt{Bi}$ for $\phi\leq\phi_J$ is commonly observed in traditional jamming transitions of dense suspensions \citep{Peters_Nature_2016,Bonn_RevModPhys2017}.
Recent findings by \cite{Rosti_NatPhys} showed that EVP fluids in inertial turbulence ($Re\gg10^3$), with subdominant elastic effects $Wi\lesssim1$, plasticity $Bi>1$ increases intermittency. We find their results for $Bi>1$ agree with our scaling $\phi \sim \sqrt{Bi}$ approaching the jamming transition (as shown in SM Fig.~S2).
Whether an analogue exponential scaling beyond the jamming transition $\phi>\phi_J$ is relevant in such inertia-dominated turbulent regimes remains to be observed.

The potential implications in practical applications of this elastic-plastic-induced jamming will be inferred by measuring the flow resistance to the power injected\citep{Berti_2008}. 
The energy injection rate per unit area (power injected),
$ \mathcal{P}_{\text{inj}} = \langle \overline{\mathbf{u} \bcdot \mathbf{F}} \rangle_{\boldsymbol{x}}
$, with $\mathbf{F}=\mathbf{F}_0+\mathbf{F}_{p}$.
In the laminar steady state, denoted by $\widehat{\cdot}$, $\partial_x \widehat{P} $ is constant and $\langle u_x \rangle_{\boldsymbol{x}} \approx \widehat{\mathrm{u}}$, where
$\widehat{\mathcal{P}_{\text{inj}}} = \frac{1}{2} V F_0 $ (see SM Sec.~S2.2). Due to the body force and the energy dissipated by the elastic-plastic-viscous effects, the flow resistance, $\mathcal{R}=\mathcal{P}_{\text{inj}}/\widehat{\mathcal{P}_{\text{inj}}} = \langle \overline{\mathbf{u} \bcdot \mathbf{F}} \rangle_{\boldsymbol{x}}/(\frac{1}{2} V F_0)$, gradually increases for $Bi\leq1$ before shifting as a consequence of the jamming phase change (\cref{fig3} \textbf{d}).
Moreover, our jamming transition scaling behaviour $\phi \sim \sqrt{Bi}$ has a potentially direct connection with other known scaling, namely, the Fanning friction factor in channel flows \citep{Rosti_JFM} and the pressure drop in porous media flows \citep{DeVita_2018}.\\
The macroscopic point of view of jamming in EVP materials is typically considered as the yield criteria, an immediate shift to a flow state once a certain stress threshold is reached, e.g., if the flow applied induces sufficiently high shear \citep{Dennin_2008,Bonn_RevModPhys2017}. 
We demonstrate the statistically homogenous dynamics of EVP Kolmogorov flows in the ET regime (dominant elastic instabilities) across a range of $Bi$  ($\phi$) align with the laminar scaling prediction below and above the transition $\phi_J$, i.e., $\phi\sim\sqrt{Bi}$ and $\phi\sim\exp{Bi}$, respectively (\cref{fig3} \textbf{e}). 
The connection to jamming in this work has profound implications on EVP fluid flows; it describes plastic events, such as plastic `plugs' in channel flows \citep{Villalba_2023,Rosti_JFM,Izbassarov_JFM_EVP_2021}, as a phase change phenomenon, whose consequences in dynamic behaviour are analogous to traditional jamming but different in the sense that it is induced by dynamic structural stress deformations arising from viscoelastic flow instabilities.
This dependency on the locality of these plastic-induced stress deformations is all important as these depend on the local geometry and flow field, implying that the nature of this jamming transition $\phi\rightarrow\phi_J$ varies from one configuration to another. For example, the square root scaling in the form $\phi \sim c_1 \sqrt{Bi} + c_2$ (see, e.g., \cref{fig3}), the constants $c_1$ and $c_2$ depend on the flow configuration, $Wi$, and $Re$. 
In support of this, our previous work \citep{DzanicFrom2024} found that two extensional flow benchmark problems, with the same dimensionless variables and similar flow type distribution, result in very different dynamic responses to plasticity with different $\phi$ across a range of $Bi$.
The consequence of this dependency, for example, is that the bulk shear rheology characterisation of the material \citep{Cheddadi_2012,Varchanis_2020}---commonly performed in Couette-type geometries---will be subject to jamming dynamics different from those experienced in the actual flow configuration of the application, making their performance in practice unpredictable. Such issues in translating rheological characterisation to quantify the flow performance in the actual flow configuration have recently been reported as major challenges in predicting the flow performance of functional EVP materials \citep{Bertsch_2022}. Predicting and controlling jamming is crucial in, e.g., 3D extrusion biomaterial printing, where the jamming of soft materials during extrusion has a negative impact on both the print quality and cell viability \citep{Xin_SciAdv_2021}.

\section{Discussion}
We have studied inertialess instabilities of elastoviscoplastic (EVP) fluid flows in the elastic turbulence regime and discovered these to transition towards jamming, featuring rich dynamics with a delicate balance between solid-like and liquid-like behaviour.
We show that in the formation of spatiotemporal narwhal coherent structures, highly localised polymer stress regions act as local low-stress energy wells, facilitating the formation of unyielded solid regions. These localised unyielded structures, in turn, deform and redistribute stresses anisotropically, leading to an interplay between viscoelastic and plastic behaviour, which dominates and absorbs more energy than viscous dissipation. 
Consequently, increasing plastic effects lead to jamming transition, sharing features directly analogous to traditional jamming of dense suspensions characterised by a drastic change in flow behaviour that is slow, minimally diffusive, and rigid-like with finite deformability leading to transverse loads.
In particular, we find the volume fraction scales as $\phi\sim \sqrt{Bi}$ until the `jamming' phase transition $\phi_J\simeq0.54$ where the behaviour changes dramatically, scaling as $\phi \sim \exp{Bi}$.
While our results are limited to inertialess flows $Re\leq1$ dominated by elasticity $Wi\gg1$, we demonstrated $\phi \sim \sqrt{Bi}$ to hold for inertia-dominated turbulent flows by \cite{Rosti_NatPhys} (SM Fig.~S2). As such, we suspect our findings to hold in shear flows where elasticity and inertia are dominant, i.e., within the elasto-inertial turbulence regime. 
The exact interactions between the flow and elastic scales with the plastic events remain unclear, and more work is required to understand the effect of elasticity $Wi$ on the scaling behaviour beyond the jamming transition $\phi > \phi_J$. 

The complex intermittent nature of inertialess EVP fluid flows makes their applications unpredictable and difficult to control, such as with the formation of plugs leading to increased flow resistance in practice. Indeed, previous works on EVP fluid flows describe the increase in volume fraction as a general effect of increasing the Bingham number $Bi$. However, no previous work has connected this phenomenon to jamming---this connection crucially reclassifies it as a phase-change phenomenon---and the analogous features which describe the impact of this transition.
The dependence of plasticity-induced stress deformations on local conditions implies that the jamming behaviour observed in one configuration may differ significantly in another. Consequently, bulk shear rheology measurements may fail to predict the jamming dynamics and flow performance of EVP materials in more application-relevant configurations ---
ranging from industrial food, cosmetic, and mining processes (e.g., moulding, extrusion, silo clogging, etc.) to emerging technologies (e.g., printing biomaterials), where unexpected jamming can lead to extreme events, disrupt performance or lead to critical failures.

\backsection[Acknowledgements]{
The authors acknowledge the assistance given by Research IT and the use of the Computational Shared Facility at The University of Manchester, and the High-Performance Computing facilities at Queensland University of Technology.}

\backsection[Data availability statement]{The data that support the findings of this study are openly available in the \textsc{JFM Notebook}, reproducing \cref{fig2,fig3}.}

\backsection[Author ORCIDs]{C. S. From, \href{https://orcid.org/0000-0001-9480-6694}{https://orcid.org/0000-0001-9480-6694}; V. Dzanic, \href{https://orcid.org/0000-0002-6691-4783}{https://orcid.org/0000-0002-6691-4783}; V. Niasar, \href{https://orcid.org/0000-0002-9472-555X}{https://orcid.org/0000-0002-9472-555X}; E. Sauret, \href{https://orcid.org/0000-0002-8322-3319}{https://orcid.org/0000-0002-8322-3319}}

\bibliographystyle{jfm}
\bibliography{Ref}

\clearpage
\bigskip
\begin{center}
\textbf{\large\textsc{Supplemental Materials}:\\
Jamming of Elastoviscoplastic fluids in Elastic Turbulence}\\
Christopher~S.~From$^{1\ast}$,
Vedad~Dzanic$^{2\dagger}$,
Vahid~Niasar$^{1}$,
Emilie~Sauret$^{2}$\\ 

    \small Corresponding authors Emails: $^\ast$christopher.from@manchester.ac.uk; $^\dagger$v2.dzanic@qut.edu.au\\
\end{center}
\setcounter{equation}{0}
\setcounter{figure}{0}
\setcounter{table}{0}
\setcounter{page}{1}
\setcounter{section}{0}
\makeatletter
\renewcommand{\theequation}{S\arabic{equation}}
\renewcommand{\thefigure}{S\arabic{figure}}
\renewcommand{\thesection}{S\arabic{section}}
\renewcommand{\thesubsection}{S\arabic{section}.\arabic{subsection}}

\section*{Supplementary Material Summary}\label{sec:Summary}
\noindent Supplementary materials consist of full details on the methods in \cref{sec:Methods} and details of the Laminar Flow solution for the theoretical calculations in \cref{sec:LaminarSolution}, containing additional results \cref{figSup1} and comparison \cref{figSup2}. Additionally, animations are provided corresponding to the results presented in Fig.~1 in the main text for $Bi=0,~0.5,~2,$ and $3$ (Supplementary Movie S1, S2, S3, and S4, respectively). 
All quantitative data presented in this work are provided, along with \textsc{JFM Notebook} providing a detailed outline of these datasets and reproduces the quantitative results in Fig.~2 and~3 in the main text.\\

\noindent
\textbf{Description of Supplementary Movie S1 to S4:} Animations are provided for the polymer conformation field ($\operatorname{tr}\mathsfbi{C}$), with the colorbar scale as in Fig.~1 and the grey areas represent the instantaneous unyielded ($\mathcal{F}=0$) solid regions within the flow field. Movie S1 to S4, respectively, corresponds to the results presented in the main text Fig.~1 for $Bi=0,~0.5,~2,$ and $3$. Captions for each Movie:\\
\textbf{Movie S1:} $\operatorname{tr}\mathsfbi{C}$ field of purely viscoplastic Kolmogorov flow in the elastic turbulence regime.\\
\textbf{Movie S2:} $\operatorname{tr}\mathsfbi{C}$ field of elastoviscoplastic Kolmogorov flow in the elastic turbulence regime at $Bi=0.5$.\\
\textbf{Movie S3:} $\operatorname{tr}\mathsfbi{C}$ field of elastoviscoplastic Kolmogorov flow in the elastic turbulence regime at $Bi=2$.\\
\textbf{Movie S4:} $\operatorname{tr}\mathsfbi{C}$ field of elastoviscoplastic Kolmogorov flow in the elastic turbulence regime at $Bi=3$.\\

\noindent
\textbf{Description of Dataset S1 to S3:} Three datasets (\textsc{NetCDF}) are provided, namely,\\ 
\textbf{Data S1:} `\texttt{supplementary\_Dataset\_VolumeFraction.nc}' \\ 
\textbf{Data S2:} `\texttt{supplementary\_Dataset\_VelProfiles.nc}'\\ 
\textbf{Data S3:} `\texttt{supplementary\_Dataset\_Other.nc}'\\ 
The \textsc{JFM Notebook} provides a detailed outline of these datasets and reproduces the quantitative
figures presented in this work.


\section{Methods}\label{sec:Methods}
\subsection{Governing Equations}
\noindent The incompressible Navier-Stokes equations in dimensionless form describe the hydrodynamic field,
\begin{equation} \label{eq:1}
    \bnabla \bcdot \mathbf{u} = 0, ~~~
    Re \frac{D\mathbf{u}}{D t} = -\bnabla P + \beta \bnabla^2 \mathbf{u} + \bnabla \bcdot \boldsymbol{\sigma} + \mathbf{F}_0.
\end{equation}
coupled with the polymer stress tensor
$$
\boldsymbol{\sigma}=\frac{1-\beta}{Wi}\Big(f \mathsfbi{C}-\mathsfbi{I}\Big),
$$ 
described by a space-time dependent conformation tensor ($\mathsfbi{C}$) constitutive equation,
\begin{equation}\label{eq:2}
    \frac{D\mathsfbi{C}}{D t} = \mathsfbi{C}\bcdot\left(\bnabla\mathbf{u}\right)
    +\left(\bnabla\mathbf{u}\right)^{\mathsf{T}} \bcdot \mathsfbi{C}
    - \frac{\mathcal{F}}{Wi}\Big(f \mathsfbi{C} - \mathsfbi{I}\Big),
\end{equation}
where $D\bullet/D t = {\partial_t \bullet} + \mathbf{u} \bcdot \bnabla  \bullet$ is the material derivative. Functions $f(s)$ and $\mathcal{F}(\sigma_v)$ are constitutive polymer models for elastic and plastic non-Newtonian behaviour, respectively. $\mathsfbi{I}$ is the identity tensor ($\operatorname{tr}\mathsfbi{I}=2$), $\mathbf{u}$ is the velocity field, and $\mathbf{F}_0$ is the external driving force (the total force is $\mathbf{F}= \mathbf{F}_p + \mathbf{F}_0$, where $\mathbf{F}_p = \bnabla \bcdot \boldsymbol{\sigma}$ is the polymer stress contribution). $Wi=\lambda V/\ell$ is the Weissenberg number, Reynolds number $Re = \rho V\ell /\mu $, and $\beta=\nu_s/\nu$ is the characteristic polymer concentration. Here, $\lambda$ is the longest polymer relaxation time, the characteristic length scale $\ell$, $\rho$ density, $V$ velocity, $\mu=\nu \rho$, and $\nu = \nu_s+\nu_p$ with $\nu_s$ and $\nu_p$ are the solvent and polymer viscosity, respectively.
For the constitutive polymer model in \cref{eq:2}, we capture a key physical behaviour of polymers using the FENE-P constitutive model, 
\begin{equation}\label{eq:FENE-P}
    f(s) = \frac{\left(L^2 - \operatorname{tr}\mathsfbi{I}\right)}{\left(L^2 - s\right)},
\end{equation} 
where $s = \mathrm{tr}\mathsfbi{C}$, imposing maximum polymer extensibility $L^2 > \mathrm{tr}\mathsfbi{C}$ that leads to shear-thinning non-Newtonian behaviour \citep{PETERLIN1961257}. We set $L^2 = 2.5\times10^3$ in all simulations.
To include the plastic effects of EVP fluids, we apply the Saramito yield-stress model \citep{Saramito_2007}, which in dimensionless form reads,
\begin{equation}\label{eq:Saramito}
   \mathcal{F}(\sigma_v)=\max\left( 0,\frac{\sigma_v-Bi}{\sigma_v} \right),
\end{equation} 
which is a function of yield stress $\sigma_v$ and the yield stress criteria $\sigma_y$ characterised by the Bingham number $Bi=\sigma_y\ell/(\mu V)$.
The yield stress is defined by $\sigma_v = \sqrt{\sigma_{J2}}$ \citep{Saramito_2007}, which in plasticity theory is known as the von Mises stress (and von Mises criterion for $\sigma_y$) \citep{Balmforth_AnnuRevFluidMech_2014}. The second invariant $\sigma_{J2} = \frac{1}{2}(\boldsymbol{\sigma}_{d}:\boldsymbol{\sigma}_{d})$ of the deviatoric part of the stress tensor $\boldsymbol{\sigma}_d = \boldsymbol{\sigma} -\mathsfbi{I}(\operatorname{tr}\boldsymbol{\sigma} /\operatorname{tr}\mathsfbi{I})$. 
The Saramito model predicts only recoverable Kelvin–Voigt viscoelastic deformation in the unyielded state ($\mathcal{F}=0$ for $\sigma_v\leq Bi$), whereas the FENE-P viscoelastic model is retained beyond yielding ($0\lneq\mathcal{F} < 1$ for $\sigma_v>Bi$). 
As in the main text, the solid (unyielded) volume fraction is calculated by 
\begin{equation}\label{eq:phi}
\phi(t)= {v_{\mathcal{F}=0}(t)}/v = \frac{\lvert\{k:\mathcal{F}(x_k,t)=0\}\rvert }{n_x n_y N^2},
\end{equation}
where $v=n_x n_y N^2$, ``:'' denotes the set formed by the unyielded regions of $k$, and ``$\lvert\cdot\rvert$'' represents the cardinality.

\subsection{Numerical Method}
\noindent Equations~\cref{eq:1} and \cref{eq:2} are solved using a symmetric positive-definite (SPD) conserving numerical solver developed in-house, comprising of the lattice Boltzmann method coupled with a high-order finite-difference scheme (see \citep{DZANIC2022105280,Dzanic_CompFluid_2022}), which was applied in our previous investigations of complex fluid flows in the ET flow regime, including purely viscoelastic instabilities \citep{vedad_jfm} and, recently, inertialess instabilities of EVP fluids [i.e., includes \cref{eq:Saramito}] \citep{DzanicFrom2024}.
Numerical considerations in \cref{eq:1,eq:2} for complex non-Newtonian fluid flow in the elastic turbulence regime ($Re\leq1$ and $El\gg 1$) are far from trivial \citep{Alves_Rev,VAITHIANATHAN20031,DzanicFrom_PRE2022,YerasiGuptaVincenzi_2023arXiv}. 
By definition, the conformation tensor is an SPD tensor $\mathsfbi{C} \gneq 0$. Conserving this property is important to prevent the rapid growth of Hadamard instabilities \citep{SURESHKUMAR199553}. SPD of $\mathsfbi{C}$ (\cref{eq:2}) is conserved by construction through the Cholesky decomposition, with the positivity enforced through the logarithmic transformation (see Ref.~\citep{VAITHIANATHAN20031}).
Moreover, the lattice Boltzmann method inherently permits exact advection for the hydrodynamic field and, as such, \cref{eq:1} is devoid of numerical diffusion. \Cref{eq:2} is solved explicitly with the high-resolution Kurganov-Tadmor scheme \citep{KT_SCHEME} applied to the advection term. Any consequent numerical diffusion arising from the numerical schemes used to solve \labelcref{eq:2} has been shown to have a negligible effect \citep{gupta_vincenzi_2019, vedad_jfm}, retrieving results in direct agreement with previous spectral studies of viscoelastic instabilities \citep{T_S_2009,DZANIC2022105280}.
Numerical regularity is added to \labelcref{eq:2} through an additional Laplacian term $ \kappa\boldsymbol{\Delta}\mathsfbi{C} $ on the right-hand-side, which effectively converts this hyperbolic equation to a parabolic form by smoothing the steep polymer stress gradients with a specified artificial diffusivity $\kappa$. While the impact of this numerical regularity strategy remains highly debated \citep{YerasiGuptaVincenzi_2023arXiv,DzanicFrom_PRE2022,Couchman_JFM_2024}, it is essential to simulate inertialess viscoelastic instabilities in the ET regime due to inherent steep polymer stress gradients at high elastic effects \citep{vedad_jfm, gupta_vincenzi_2019}. We minimize $\kappa$ and any associated artefacts by setting the admissible Schmidt number $Sc=\nu_s/\kappa=10^3$ as in \citep{Berti_2008}, a typical value used across numerical methods, including, e.g., spectral methods \citep{T_S_2009,Buza_JFM_2022}.

\subsection{Viscoelastic Kolmogorov Flow and periodic boundary conditions}
The viscoelastic Kolmogorov flow problem is solved in a 2D domain $\boldsymbol{x}$ with double periodic boundary conditions (PBCs) where a single unit cell is $\left[0,2\pi\right]^2$. Let $\boldsymbol{n}$ be the level of periodicity in each direction, for which $n_x,n_y>1$ results in $n_x\times n_y$ unit cells to be solved.
Based on our previous investigation \citep{vedad_jfm,DzanicFrom2024}, including the numerical regularity term $ \kappa\boldsymbol{\Delta}\mathsfbi{C} $ in \labelcref{eq:2} with a single unit cell, i.e., $\boldsymbol{n}=\boldsymbol{1}$, will result in unphysical artefacts arising from insufficient periodicity and, in turn, inaccurate simulation of ET flow regimes. 
For all simulations, we set $n_x=6$ and $n_y=4$, ensuring unicity is conserved, with $N^2=128^2$ grid points in each unit cell, imposing a spatial resolution of $2\pi/N$ and a total of $n_x N \times n_y N= 768 \times 512 $ grid points. 
The Kolmogorov shear flow is driven by a constant external forcing $\boldsymbol{F}$ in \cref{eq:1}, defined as 
\begin{equation}\label{eq:KF}
   F_x(y) = F_0 \cos(K y), ~ F_y = 0.
\end{equation}
where the amplitude $F_0=V\nu K^2$ and $K=2$ is the spatial frequency $\ell=1/K$ and $y/\ell=[0,n_y\times 2 \pi]$.
To simulate ET, a small perturbation $\boldsymbol{\delta}$, a random Gaussian (normally distributed) noise in the order of $\mathcal{O}(V\times 10^{-3})$, is introduced to the initial velocity field $\boldsymbol{u}_{t=0}+\boldsymbol{\delta}$ to induce viscoelastic instabilities.
In all simulations, dimensionless groups are set following previous numerical investigations \citep{Berti_2008, DzanicFrom2024,vedad_jfm}, including $\beta=0.9$, $Wi=20$, $L^2 = 2.5\times10^3$ in \cref{eq:FENE-P}, artificial diffusivity $Sc=10^3$, and set $Re=1$ below the critical value at which inertial instabilities arise $Re<\sqrt{2}$ \citep{Boffetta_PRE_2005}.

\section{Laminar Flow Solution}\label{sec:LaminarSolution}

\noindent The laminar flow solution of the governing equations [\cref{eq:1,eq:2}] is obtained at equilibrium, i.e., under the assumption that the flow is statistically homogeneous, such that 
\begin{equation}\label{eq:A1}
    \frac{\partial \mathbf{u}}{\partial t} = 0 \quad \text{and} \quad \frac{\partial \mathsfbi{C}}{\partial t} = 0, \quad t\geq\overline{t}, 
\end{equation}
where the velocity field is immediately defined from the constant driving force in \labelcref{eq:1}. Specifically, in Kolmogorov flow \labelcref{eq:KF}, $\mathrm{F}_y = 0$, imposing a constant pressure gradient such that the velocity field in the laminar solution simplifies to a constant one-dimensional variation
\begin{equation}\label{eq:A2}
   u_x(y) = V \cos(K y), \quad u_y = 0,
\end{equation}
with all gradients in the $y$-direction, i.e.,
\begin{equation}\label{eq:A3}
    \bnabla \mathbf{u} = \begin{pmatrix}
   0 & \partial_y u_x \\
   0 & 0
   \end{pmatrix} = \begin{pmatrix}
   0 & - K V \sin(K y) \\
   0 & 0
   \end{pmatrix}.
\end{equation}
The inertial term and the elastic stress transport term of the corresponding material derivatives $\mathbf{u} \bcdot \bnabla  \bullet$ are zero.
Substituting \cref{eq:A1,eq:A2,eq:A3} into the governing equation for polymer conformation tensor \labelcref{eq:2}, 

\begin{align}
\label{eq:lam_Cxx0}
    \frac{\mathcal{F}}{Wi} \left( f(s)\mathsfit{C}_{xx} - 1 \right) &= 2 \mathsfit{C}_{xy} \partial_y u_x,\\
\label{eq:lam_Cyy0}
    \frac{\mathcal{F}}{Wi}\left( f(s)\mathsfit{C}_{yy} - 1 \right) &= 0,\\
\label{eq:lam_Cxy0}
    \frac{\mathcal{F} f(s)}{Wi}\mathsfit{C}_{xy} &= \partial_y u_x, \quad \mathsfit{C}_{xy} = \mathsfit{C}_{yx},
\end{align}

where $\kappa$ is neglected such that the numerical regularity term is $\kappa\boldsymbol{\Delta}\mathsfbi{C}=\boldsymbol{0}$, since the base laminar solution is independent of this term for small $\kappa\ll10^{-3}$. 
\cref{eq:lam_Cyy0,eq:lam_Cxy0} are immediately obtained,

\begin{equation}\label{eq:lam_Cyy}
    \mathsfit{C}_{yy} = 1,
\end{equation}
and
\begin{equation}\label{eq:lam_Cxy}
    \mathsfit{C}_{xy} = \frac{Wi}{\mathcal{F} f(s)}\partial_y u_x 
    = K\frac{Wi}{\mathcal{F} f(s)}\operatorname{sin}(Ky).
\end{equation}
Substituting \labelcref{eq:lam_Cxy} into \labelcref{eq:lam_Cxx0},
\begin{equation} \label{eq:lam_Cxx}
    \mathsfit{C}_{xx} = f^{-1}(s)\Big(1 + \frac{2 Wi^2}{\mathcal{F}^2 f(s)}  (\partial_y u_x)^2\Big) = f^{-1}(s)\Big(1 + \frac{2 K^2Wi^2}{\mathcal{F}^2 f(s)} \operatorname{sin}^2(Ky)\Big),
\end{equation}
as in the main text Eq.~(3.3). The polymer stress tensor is then calculated with 
$
\boldsymbol{\sigma}=(1-\beta) \Big( Wi^{-1}\big(f(s)\mathsfbi{C}-\mathsfbi{I}\big)\Big).
$

\Cref{eq:lam_Cxx0,eq:lam_Cyy0,eq:lam_Cxy0} are well-known solutions for purely viscoelastic fluids using the Oldroyd-B model (see, e.g.,~\citep{Boffetta_2005,Berti_2010,Gorodtsov_1967}) and for other constitutive models, e.g., the FENE-P \citep{Couchman_JFM_2024}. Here, we include, in addition, $\mathcal{F}$ \cref{eq:Saramito} based on the Saramito yield-stress model for the plastic contribution of an EVP fluid.
We theoretically approximate $\phi$ \labelcref{eq:phi} by solving for the yield stress based on laminar solution [\labelcref{eq:lam_Cyy,eq:lam_Cxx,eq:lam_Cxy}] under the assumption that in the statistically stationary regime in the Kolmogorov flow configuration, the flow and polymer stress fields are accurately described by sinusoidal profiles \citep{Berti_2010,Boffetta_PRE_2005}. Here, the finite extensibility function $f(\operatorname{tr}\mathsfbi{C})$ \labelcref{eq:FENE-P} is obtained directly from the initial stress field imposed by the applied flow field \labelcref{eq:A3}, i.e., $\operatorname{tr}\mathsfbi{C} \cong 2 + 2 K^2 Wi^2 \operatorname{sin}^2(Ky) $. The stress field $\mathsfbi{C}$ \labelcref{eq:lam_Cyy,eq:lam_Cxx,eq:lam_Cxy} and $\mathcal{F}$ \labelcref{eq:Saramito} are then solved iteratively.
%
%
\begin{figure}
	\centering
	\includegraphics[width=0.50\columnwidth]{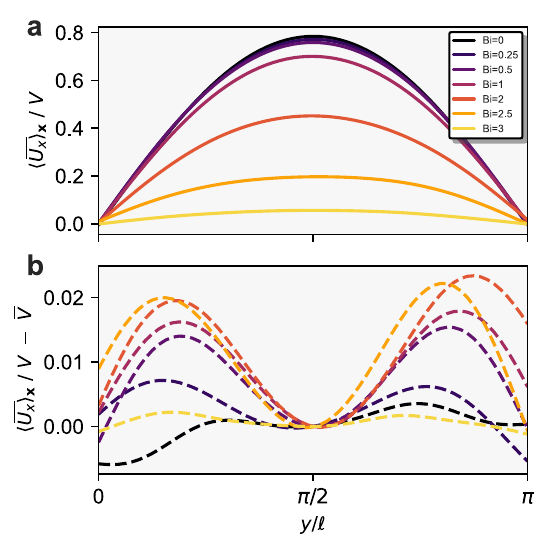}
	\caption{Velocity profile deviation from the base sinusoidal profile with increasing $Bi$. (\textbf{\textsf a}) Spatiotemporal mean velocity profiles $\langle \overline{U_x} (y)\rangle$ and (\textbf{\textsf b}) their deviation from the base sinusoidal profile scaled by the amplitude of each $Bi$ case, i.e., $\overline{V}(y)=\max_y(\langle \overline{U_x} (y)\rangle)\operatorname{cos}(Ky)$.}
	\label{figSup1}
\end{figure}
The laminar solution predicts the volume fraction $\phi$ \labelcref{eq:phi} due to the applied stress, neglecting the dynamic deformation of the stress field due to the unyielded regions. 
It is expected to be valid in the case of Kolmogorov flow configuration for cases where the statistically steady mean streamwise flow profiles (\cref{figSup1} \textbf{a}) retain the sinusoidal profile \labelcref{eq:A2}. We obtain the mean streamise flow profile by averaging all $\mathrm{u}_x$ profiles in $[0,n_y \times 2\pi]$, i.e., 
$$
\langle \overline{U_x} (y^*)\rangle = (K n_y)^{-1}\sum_{k=0}^{K n_y-1}\lvert \langle \overline{u_x}(y^* + k\pi)\rangle_{{x}} \rvert,
$$
along each profile $y^* \in [0,\pi)$, as in Fig.~3 \textbf{e} in the main text, in which the base sinusoidal profile, i.e., $\operatorname{cos}(Ky)$, scaled by the amplitude $\max_y(\langle \overline{U_x} (y)\rangle)$ of each $Bi$ case is superimposed to measure the deviation from a sinusoidal profile.
The departure from this sinusoidal profile, that is, $\overline{V}(y)=\max_y(\langle \overline{U_x} (y)\rangle)\operatorname{cos}(Ky)$, is shown quantitatively for each $Bi$ case in \cref{figSup1} \textbf{b}, and appears to increase with fluctuations in $\phi$ (see Fig.~2 \textbf{b} and \textbf{c} in the main text). We suspect that this is due to the fact that most of the domain behaves as a viscoelastic solid at $Bi>2$ (where $\phi>0.5$) or a viscoelastic fluid at $Bi<2$ (where $\phi<0.5$), with the maximum spatiotemporal distribution of yielded and unyielded regions at $Bi=2$ (at $\phi_J,~ \phi\sim0.5$). Hence, deviation from the based sinusoidal profile in \cref{figSup1} \textbf{\textsf{b}} increases with $1-|(\phi-0.5)/0.5|$.
As shown in the main text in Fig.~3 \textbf{e}, indeed, the laminar solution predicts $\phi$ for $Bi<2.5$.
While the deviation from the sinusoidal profile is minimum at $Bi=3$ (\cref{figSup1} \textbf{\textsf{b}}), the numerical solution exceeds the theoretically approximated $\phi$ (Fig.~3 \textbf{e}) as the shear regions in the laminar solution are constant. In fact, as discussed in the main text, prior to transitioning to the jammed state for $Bi=3$, in the initial steady-state region $300T\lesssim t\lesssim800T$, the numerical solution is consistent with the theoretically approximated $\phi$ (Fig.~2 \textbf{a}).

\subsection{Flow Resistance}
\noindent In the laminar solution, the power injected is,
\begin{equation}
    \int_\Omega \widehat{\mathbf{u}} \bcdot \mathbf{F} \, dA = \int_0^{\ell} \int_0^{\ell} \widehat{\mathbf{u}} \bcdot \mathbf{F} \, dx \, dy ,
    \label{eq:InjectPressure}
\end{equation}
   
   where $\Omega = [0, \ell]^2$ is the periodic domain, i.e., a single unit cell $\ell=\ell_x=\ell_y$.
   Solving \cref{eq:InjectPressure} above for the Kolmogorov flow, where the constant body force $\mathbf{F}_0(x, y) = (F_0 \cos(y/\ell_y),~0)$ and at laminar steady state $\widehat{\mathbf{u}}=(V\cos(y/\ell_y),~0)$, such that $\widehat{\mathbf{u}} \bcdot \mathbf{F}_0 = (V F_0 \cos^2(y/\ell_y),~0)$, gives, 
   $$
   \int_0^{\ell} \int_0^{\ell} \widehat{\mathbf{u}} \bcdot \mathbf{F} \, dx \, dy \approx \frac{1}{2} V F_0  \ell^2. 
   $$
Similarly, the power injected per unit area is,
\begin{equation}
    \widehat{\mathcal{P}_{\text{inj}}} = \frac{1}{\ell^2}\int_0^{\ell} \int_0^{\ell} \widehat{\mathrm{u}} \bcdot \mathbf{F} \, dx \, dy \approx \frac{1}{2} V F_0,
    \label{eq:InjectPressure_Area}
\end{equation}
   such that the flow resistance $\mathcal{R} = \mathcal{P}_{\text{inj}} / \widehat{\mathcal{P}_{\text{inj}}}$, where $ \mathcal{P}_{\text{inj}} = \langle \overline{\mathbf{u} \bcdot \mathbf{F}} \rangle_{\boldsymbol{x}}$ (with $\mathbf{F}=\mathbf{F}_0+\mathbf{F}_{p}$) is the power injected per unit area measured from the numerical solution in the statistically homogeneous region, is calculated by,
   $$
   \mathcal{R}= \frac{\langle \overline{\mathbf{u} \bcdot \mathbf{F}} \rangle_{\boldsymbol{x}}}{\frac{1}{2} V F_0},
   $$
   as in the main text.
   
\subsection{$\phi\sim\sqrt{Bi}$ scaling}\label{SM:Supplementary}
\begin{figure}
	\centering
	\includegraphics[width=0.35\columnwidth]{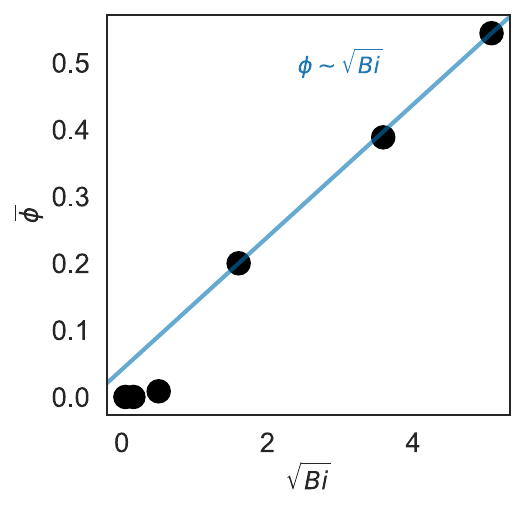}
	\caption{ Volume fraction scaling $\phi\sim\sqrt{Bi}$ for EVP turbulent flows with subdominant elasticity ($Wi=1$) by Ref.~\citep{Rosti_NatPhys}.}
	\label{figSup2}
\end{figure}

In \cref{figSup2}, we demonstrate our scaling $\phi\sim\sqrt{Bi}$ to hold for EVP fluids in inertially driven, turbulent flows ($Re\gg100$) with subdominant elasticity ($Wi=1$), recent work by Abdelgawad~\textit{et al.} \citep{Rosti_NatPhys} (data available therein).

\end{document}